\documentclass[aps,pra,reprint,superscriptaddress,longbibliography]{revtex4-1}
\usepackage{amsmath,amssymb,graphicx,bm,epstopdf}

\begin{document}

\title{Temporal dynamics of light-written waveguides in unbiased liquid crystals}

\author{Alessandro Alberucci}
\affiliation{Institute of Applied Physics, Abbe Center of Photonics, Friedrich Schiller University Jena, Albert-Einstein-Stra{\ss}e 15, 07745 Jena, Germany}
\email{alessandro.alberucci@gmail.com}

\author{Raouf Barboza}
\affiliation{Photonics Laboratory, Tampere University of Technology, FI-33101 Tampere, Finland}
\affiliation{Dipartimento di Fisica ``Ettore Pancini", Universit\`{a} di Napoli Federico II, Complesso Universitario di Monte Sant'Angelo, IT-80126 Napoli, Italy}

\author{Chandroth P. Jisha}
\affiliation{Centro de F\'{\i}sica do Porto, Faculdade de Ci\^encias, Universidade do Porto, 
Porto 4169-007, Portugal}

\author{Stefan Nolte}
\affiliation{Institute of Applied Physics, Abbe Center of Photonics, Friedrich Schiller University Jena, Albert-Einstein-Stra{\ss}e 15, 07745 Jena, Germany}
\affiliation{Fraunhofer Institute for Applied Optics and Precision Engineering, Albert-Einstein-Straße 7, 07745 Jena, Germany}



\begin{abstract}

The control of light by light is one of the main aims in modern photonics. In this context, a fundamental cornerstone is the realization of light-written waveguides in real time,
resulting in all-optical reconfigurability of communication networks. Light-written waveguides are often associated with spatial solitons, that is, non-diffracting waves due to a
nonlinear self-focusing effect in the harmonic regime.
From an applicative point of view, it is important to establish the temporal dynamics for the formation of such light-written guides. Here we investigate theoretically the temporal dynamics in nematic liquid crystals, a material where spatial solitons can be induced using continuous wave (CW) lasers with few milliWatts power.
We fully address the role of the spatial walk-off and the longitudinal nonlocality in the waveguide formation. We show that, for powers large enough to induce light self-steering, the beam undergoes several fluctuations before reaching the stationary regime, in turn leading to a much longer formation time for the light-written waveguide.

\end{abstract}





\maketitle


\section{Introduction}

Nowadays, in optical communication systems light beams carry the information via amplitude or phase modulations originating from an electric signal, either a voltage or a current.
The signal processing is usually made by electronic circuits: the consequent transduction from optical to electrical signal, back and forth, represents today the most important
limitation on the maximum achievable speed in optical communications. Accordingly, a big effort has been dedicated to realize all-optical systems, where direct optical signal
processing and light-controlling-light schemes can be realized \cite{Willner:2014}. All-optical signal processing typically (see Refs.~\cite{Zhang:2012,Zhao:2016} for
counterexamples) implies a nonlinear response in the material, that is, changes in the optical properties of the medium due to light \cite{Willner:2014}. \\
Although in some cases freely propagating beams are used to transmit information \cite{Chan:2006}, typically the optical beams are localized in a specific spatial region using
waveguides, the most famous being fiber optics \cite{Yariv:1997}. Optical waveguides allow to minimize the footprint of photonic devices, one of the most important advantages of
integrated optics with respect to free space optics. Starting from the concepts of all-optical signal processing and integrated optics, the next breakthrough is to realize guiding
structures defined by light in real time, paving the way to all-optical control of the geometry and topology of optical networks. The natural candidates for this are
spatial solitons (SSs), waves subject to spatial localization owing to the action of nonlinear self-focusing \cite{Bjorkholm:1974,Stegeman:1999,Kivshar:2003}. In fact, spatial
solitons are usually associated with a light-written waveguide inside the material \cite{Kelly:1965,Snyder:1991,Duree:1993,Rotschild:2005,Weining:2013,Kelly:2016,Bezryadina:2017,Yadira:2017}, capable to
guide as well as probe beams at other wavelengths and low power. \\
One of the most used material for the generation of stable two-dimensional SSs is nematic liquid crystals (NLCs) \cite{Peccianti:2012,Assanto:2012}. In NLCs several different
mechanisms can lead to nonlinear optical effects, according to the NLC composition, the optical excitation(s) and the environmental conditions \cite{Simoni:1997}. For CW inputs, the strongest contribution usually comes from reorientational nonlinearity, consisting in a collective rotation of the molecules when subject to an optical torque \cite{Tabyrian:1980}.
Reorientational nonlinearity in NLC is very large, allowing the observation of nonlinear effects with input powers of few milliWatts. On the other side, the time response is quite
slow, ranging from milliseconds to seconds, according to the geometry and size of the NLC cell \cite{Simoni:1997}. The effect of self-focusing induced by
reorientational nonlinearity on long propagation distances (i.e., beyond the plane wave approximation) has been first studied by Braun in 1993 \cite{Braun:1993}. In the latter
experiment, the presence of the Fr\'eedericksz transition (FT) inhibited the observation of a stable SS. Stable SSs in NLCs have been found in 2000 applying a bias to the NLC cell
in order to change the initial direction of the NLC molecules and avoid the appearance of the FT \cite{Peccianti:2000}. Later on, such approach has been generalized to unbiased
cells by changing the anchoring conditions of the NLC molecules at the edge of the sample \cite{Peccianti:2004}. Unlike Townes soliton (i.e., SS in a local Kerr media
\cite{Chiao:1964}), SS in NLCs, often called nematicons \cite{Peccianti:2012}, are stable due to the spatially nonlocal nonlinear response \cite{Suter:1993,Bang:2002,Conti:2003}
stemming from the strong elastic interaction between adjacent molecules \cite{Peccianti:2012}. \\
Here we are interested in the investigation of the temporal response in the formation of nematicons in unbiased NLC cells \cite{Peccianti:2004}. From an applicative point of view,
the knowledge of the dynamical behavior of nematicons is fundamental to understand the maximum speed achievable in the modification of the guiding structure. The dynamics of
nematicon formation has been already studied in Ref.~\cite{Beeckman:2005,Strinic:2005} in the case of a biased cell. The absence of bias makes the self-written waveguides
qualitatively different. As a matter of fact, the static (or low-frequency \cite{Peccianti:2000}) electric field creates a transverse gradient in the refractive index distribution,
the maximum being at the center of the cell. The bias-induced inhomogeneity works both as a strongly multimodal waveguide \cite{Beeckman:2010} and a potential landscape where
the soliton starts to oscillate \cite{Peccianti:2004_1,Beeckman:2005_1}. Such dynamics is strongly affected by spatial walk-off \cite{Peccianti:2004} as well, a fundamental
ingredient in the propagation of nematicons \cite{Braun:1993,Alberucci:2010_2} neglected in Ref.~\cite{Beeckman:2005}.

\begin{figure}[htbp]
\centering
\includegraphics[width=\columnwidth]{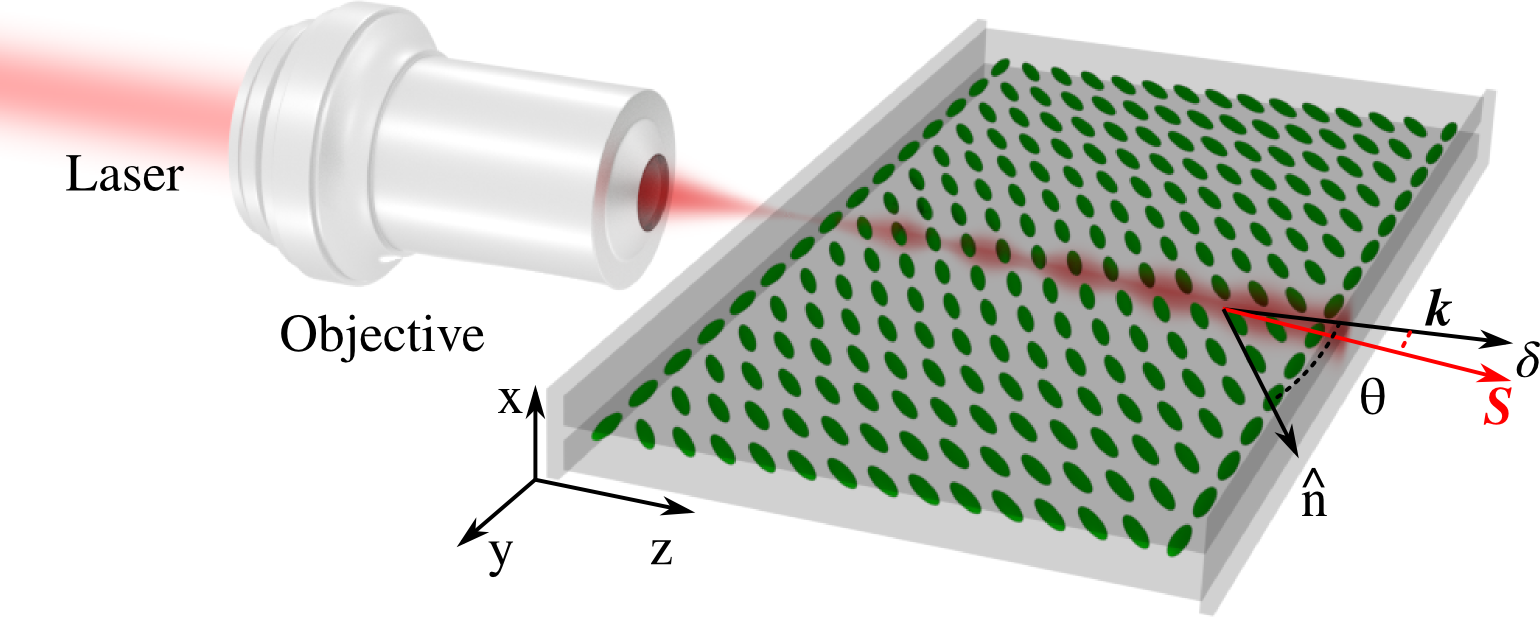}
\caption{Sketch of the studied geometry. Input beam is impinging normal to the input interface (wavevector $\bm{k}$ parallel to the axis $z$). The Poynting vector $\bm{S}$ forms an angle $\delta$ with $\bm{k}$ due to the medium anisotropy. The NLC molecules are represented by the green ellipsoids. The long axis of each ellipsoid correspond to the local orientation of the director $\hat{n}$. }
\label{fig:cell_sketch}
\end{figure}

\section{Geometry and model}

We consider a planar cell of thickness $L_x=100~\mu$m, length $L_z$ and infinitely extended along $y$. 
Two glass slides, parallel to the plane $yz$, provide the NLC confinement. The two slides are treated in order to provide a uniform alignment of the NLC molecules in the absence of
external excitations, the direction in the bulk being at $\theta_0=45^\circ$ with respect to $\hat{z}$. We assume strong anchoring conditions, that is, the molecular alignment in proximity of the NLC/glass interface is not affected by the impinging optical field \cite{Simoni:1997,Peccianti:2012}. 
A fundamental Gaussian beam at $\lambda=$1064 nm is injected into the sample with wavevector $\bm{k}=k_0\hat{z}=\frac{2\pi}{\lambda}\hat{z}$ parallel to $\hat{z}$. 
Two other very thin glass slides, inducing a molecular alignment parallel to $\hat{y}$, are placed at the input ($z=0$) and the output ($z=L_z$) interfaces to avoid the meniscus that forms at NLC/air interface. The interfaces induce a perturbation on the molecular distribution on distances $\approx L_x$, which we will neglect in this work. Noteworthy, in the paraxial limit the perturbation close to the input interface does affect neither the input polarization nor the input wavevector, the main effect being a slight transverse shift owed to the $z-$dependent walk-off angle. \\
The mean molecular direction is provided by a vector field, called the director $\hat{n}$, and represented by the long axis of the green ellipses in Fig.~\ref{fig:cell_sketch}. Here we are interested in reorientational nonlinearity, that is, rotation of the director induced by the optical field \cite{Simoni:1997,Assanto:2012}. Thus, we will focus on $y$-polarized inputs ($x$-polarization induces reorientation only above a power threshold \cite{Durbin:1981}). Due to the NLC anisotropy, the beam inside the cell propagates (i.e., the Poynting vector $\bm{S}$) forming a walk-off angle $\delta$ with $\hat{z}$. Hereafter we will also neglect effects associated with changes in temperature (in this paper we focus our attention on undoped NLC away from the isotropic-nematic transition; we also assume that the saturation of the reorientational nonlinearity is not achieved) \cite{Assanto:2012,Alberucci:2018}, considering the latter constant in time and space. Due to the chosen input polarization, the director is constrained to lie on the plane $yz$, thus the angle $\theta$ formed with respect to $\hat{z}$ describes univocally $\hat{n}$. Optically, the system behaves like an inhomogeneous positive uniaxial material, with refractive indices equal to $n_\bot$ and $n_\|$ for electric fields oscillating normal and parallel to $\hat{n}$, respectively. The extraordinary waves experience a refractive index $n_e(\theta)=\left(\cos^2\theta/n^2_\bot + \sin^2\theta/n^2_\| \right)^{-1/2}$. We also define the optical anisotropy $\epsilon_a=n_\|^2-n_\bot^2$. \\
After setting $H_x=Ae^{ik_0 n_e(\theta_b)z}$, the nonlinear light propagation in the paraxial approximation ($\partial^2_zA=0$) and in the harmonic regime (electromagnetic field
$\propto e^{-i\omega t}$) can be described by \cite{Alberucci:2010_2}
\begin{align}
  & 2i k_0 n_e(\theta_b) \left(\frac{\partial A}{\partial z} +\tan\delta(\theta_b) \frac{\partial A}{\partial y} \right) + \frac{\partial ^2 A}{\partial x^{ 2}} \nonumber \\  & + D_y \frac{\partial ^2 A}{\partial y^{ 2}} + k_0^2 \Delta n_e^2(\theta) A=0. \label{eq:NLSE}
\end{align}
In Eq.~(\ref{eq:NLSE}) we introduced the nonlinear index well $\Delta n_e^2(\theta)=n_e^2(\theta)-n_e^2(\theta_b)$ and the diffraction coefficient
$D_y=n_e^2(\theta_b)/\left(\epsilon_\bot +\epsilon_a \cos^2 \theta_b \right)$, where $\theta_b(z)$ is the average angle perceived by the beam on each section $z=constant$.
Equation~(\ref{eq:NLSE}) accounts for nonperturbative nonlinear effects, including self-steering and nonlinearity saturation \cite{Alberucci:2010_3}. \\
In writing Eq.~(\ref{eq:NLSE}) we assumed that the medium is stationary, that is, the reorientation angle $\theta$ does not depend on time $t$. In our case this hypothesis clearly does
not hold valid. In fact, the reorientation of the NLC molecules, under the single elastic constant approximation and neglecting inertial effects in the NLC dynamics \cite{Au:1991},
is governed by
\begin{align}
   & \nu \frac{\partial \theta}{\partial t} = K\nabla^2\theta + \frac{\epsilon_0 \epsilon_a}{4} \sin\left[2\left(\theta-\delta_b  \right) \right] \left( |\bm{E}_t|^2 - |\bm{E}_s|^2 \right) \nonumber \\ &  \ \ \ \ \ \  + \frac{\epsilon_0 \epsilon_a}{2} \cos\left[2\left(\theta-\delta_b \right) \right] \text{Re}\left({E}_t {E}_s^* \right),
    \label{eq:reorientation}
\end{align}
where we defined $\delta_b=\delta(\theta_b)$, $K$ is the Frank elastic constant, and $\nu$ is the rotational viscosity. The subscripts $t$ and $s$ indicate the component of the
electric field parallel to $\hat{t}(z)=\hat{y}\cos\delta_b(z) - \hat{z} \sin\delta_b(z)$ (direction of the electric field for the plane wave with $\bm{k}\|\hat{z}$) and $\hat{s}(z)=\hat{y}\sin\delta_b(z)+\hat{z}\cos\delta_b(z)$ (direction of the Poynting vector $\bm{S}$), respectively. Equation~(\ref{eq:reorientation}) thus provides a time-dependent $\theta$, seemingly in
disagreement with the assumptions behind Eq.~(\ref{eq:NLSE}). Nonetheless, the time response of the optical field to material changes is much faster than typical NLC response
times, thus we can safely suppose a quasi-stationary regime for the electromagnetic field, since the latter adapts instantaneously to the refractive index landscape. \\
Equation~(\ref{eq:reorientation}) must be solved jointly with the boundary conditions $\theta=\theta_0$ at all the edges (i.e., $x=0$, $x=L_x$, $|y|\rightarrow\infty$, $z=0$ and
$z=L_z$) of the NLC layer. For the sake of simplicity, we introduce the nonlinear perturbation on the director angle $\psi=\theta-\theta_0$, corresponding to a vanishing condition
at the boundaries.\\
Let us first discuss qualitatively solutions of Eq.~(\ref{eq:reorientation}) for a fixed optical field with a spatial extension smaller than $L_x$ (illumination starting at $t=t_0$)
with $\bm{E}_s=0$. In the stationary regime (i.e., $t\rightarrow\infty$), it is well known that Eq.~(\ref{eq:reorientation}) provides a nonlinear perturbation extending on an area
of transverse size comparable with $L_x$, regardless of the input beam width $w_{in}$ \cite{Khoo:1987,Alberucci:2007}. If $w_{in}\ll L_x$, nonlinear light propagation is in the
highly nonlocal regime. However, during the transient, the behavior of $\psi$ is different. Specifically, for short time intervals after the illumination, in Eq.~(\ref{eq:reorientation}) the elastic forces can be neglected, providing (see also Eq.~(\ref{eq:transf_short_time}) in the Appendix)
\begin{equation}  \label{eq:reorientation_short_time}
  \theta(t)=\theta_0 + \frac{\epsilon_0\epsilon_a}{4\nu}\sin\left[2 (\theta_0 -\delta_0) \right] |\bm{E}_t|^2 (t-t_0),
\end{equation}
where $\delta_0=\delta(\theta_0)$ and where we supposed small perturbations, that is, $\psi\ll \theta_0$. According to Eq.~(\ref{eq:reorientation_short_time}), in this regime the
NLC behaves spatially like a local material \cite{Henninot:2002,Warenghem:2006}, whereas it shows a highly nonlocal response in the time domain \cite{Conti:2010_1}. Physically, the
optically-induced perturbation has not enough time to spread away from the optical beam due to the slow response time of the intermolecular elastic forces
\cite{Beeckman:2005,Henninot:2006}. Later on, the perturbation recovers its nonlocal nature, thus providing a stable optical soliton. \\
To evaluate the interval of validity $\Delta t$ for Eq.~(\ref{eq:reorientation_short_time}), we can substitute Eq.~(\ref{eq:reorientation_short_time}) into the RHS of
Eq.~(\ref{eq:reorientation}) and find when the optical and the elastic torque are equal to each other in absolute value. We find
\begin{equation}  \label{eq:validity_short_time}
  \Delta t= \frac{\nu}{K}\frac{|\bm{E}_t|^2}{\left|\nabla^2|\bm{E}_t|^2\right|}.
\end{equation}

According to Eq.~(\ref{eq:validity_short_time}), $\Delta t$ depends on the NLC material (parameter $\nu/K$, here the diffusion coefficient)  and on the shape of the optical
illumination, but not on the incident optical power. In particular, the larger the spatial derivatives of the intensity distribution the shorter is the time required for the
appearance of spatially nonlocal effects.

\section{Solution in a longitudinally-invariant geometry}
\label{sec:z_invariant}
To verify the validity of Eq.~(\ref{eq:validity_short_time}), we consider the simplest case of a structure invariant along $z$ \cite{Au:1991,Veltri:2007}.
Hereafter, we will consider the NLC mixture E7 featuring $\nu=0.2$ Pa~s and $K=12$ pN. At $\lambda=1064~$nm the E7 refractive indices are approximately $n_\|=1.7$ and $n_\bot=1.5$.
For the optical illumination we make the ansatz $|\bm{E}_t|^2 = \frac{4Z_0 P}{\pi n_e w^2} \exp{(-2r^2/w^2)}$ (we set $r=\sqrt{x^2+y^2}$), yielding $\Delta t=  \nu w^2 \left/ \left(16 K \left| \frac{r^2}{w^2}- \frac{1}{2} \right| \right) \right.$. Given that the transition time changes across the beam cross-section, we need to consider the minimum of $\Delta t$ on the transverse plane $xy$, the latter being achieved when $r\rightarrow\infty$, with the constraint that the field should not be negligibly small. Taking conventionally the point $r=2.35w$ (such a choice comes from the direct comparison with exact numerical simulations, see below), we find
\begin{equation}  \label{eq:min_time}
  \min(\Delta t) \approx \frac{\nu}{K} \frac{w^2}{80}.
\end{equation}
Thus, Eq.~(\ref{eq:reorientation_short_time}) holds valid on the temporal interval $[t_0,\ t_0+T]$, where we set $T=\min(\Delta t)$.

\begin{figure}[htbp]
\centering
\includegraphics[width=0.47\textwidth]{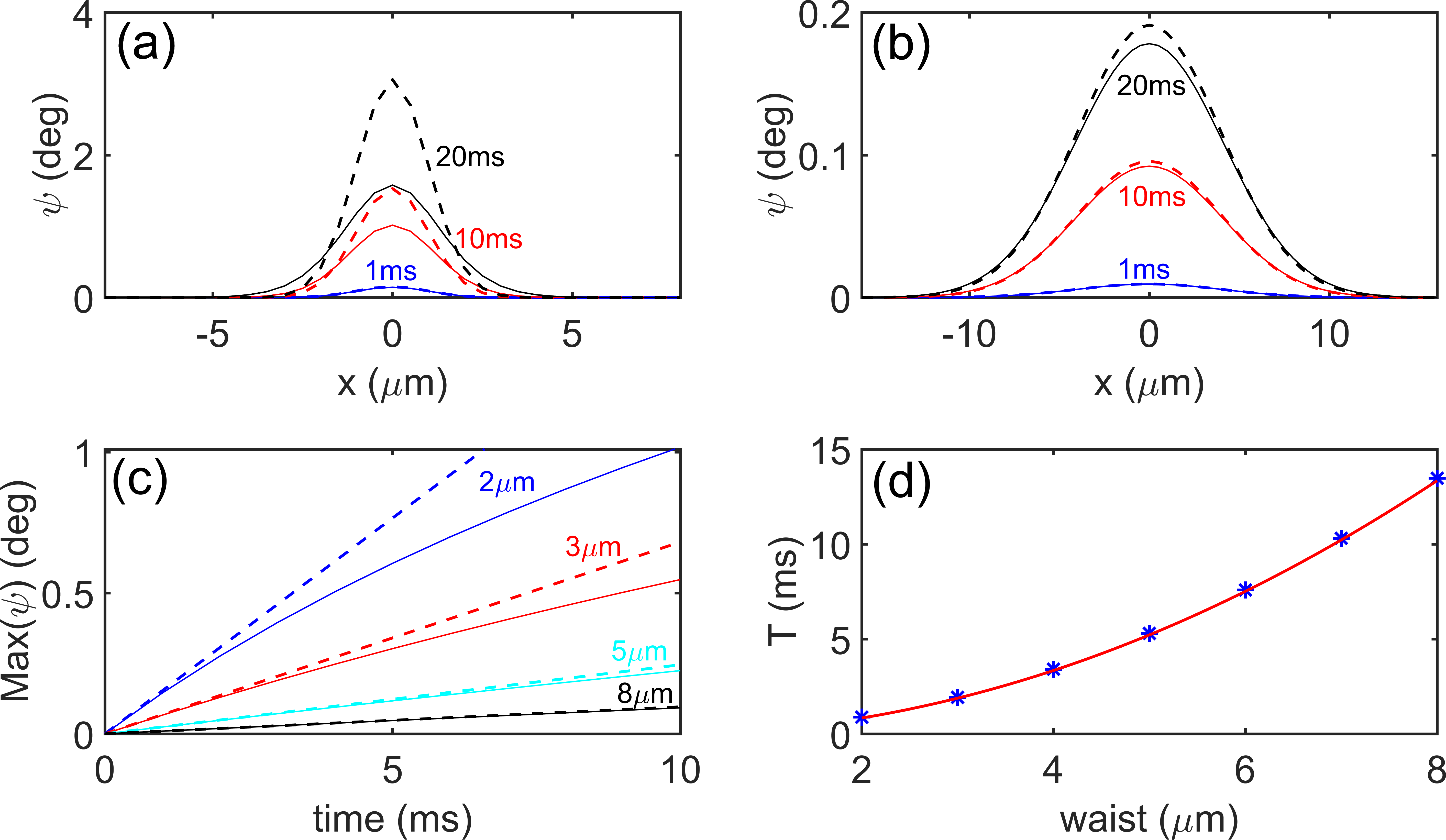}
\caption{Optical perturbation $\psi(x,y=0)$ versus $x$ at different times for a Gaussian beam of waist (a) $w=2~\mu$m and (b) $w=8~\mu$m. Solid and dashed lines correspond to exact
numerical simulations and to the approximated formula Eq.~(\ref{eq:reorientation_short_time}), respectively. (c) Behavior of the maximum perturbation angle $\psi_m$ versus time for
numerical simulations (solid lines) and short-time approximation (dashed lines), for four different beam widths. (d) Time interval $T$ versus the beam waist $w$ according to the
numerical simulations [symbols; $T$ corresponds to a difference in $\psi_m$ of 5$\%$ with respect to Eq.~(\ref{eq:reorientation_short_time})] and theoretical prediction
Eq.~(\ref{eq:min_time}). The power is fixed to 5 mW and $\theta_0=\pi/4$. Here walk-off is neglected for the sake of simplicity.}
\label{fig:short_time}
\end{figure}
Thus, according to Eq.~(\ref{eq:min_time}), the narrower the beam the shorter is the validity range of Eq.~(\ref{eq:reorientation_short_time}). For example, for $w=2~\mu$m
Eq.~(\ref{eq:min_time}) provides $\min(\Delta t) \approx$ 1 ms.\\
Figure~\ref{fig:short_time}(a-b) compare the all-optical perturbation $\psi(x,y,t)$ given by Eq.~(\ref{eq:reorientation_short_time}) (dashed lines) and exact solutions (solid lines)
computed via numerical simulations of Eq.~(\ref{eq:reorientation}). For quantitative comparison, we take the maximum of $\psi(x,y,t)$ at a fixed instant, let us call it $\psi_m(t)$.
As predicted, for short times values for $\psi_m$ provided by the exact solution of Eq.~(\ref{eq:reorientation}) and approximate solution from
Eq.~(\ref{eq:reorientation_short_time}) perfectly overlap [Fig.~\ref{fig:short_time}(c)], on a temporal interval $T$ in good agreement with Eq.~(\ref{eq:min_time})
[Fig.~\ref{fig:short_time}(d)]. Numerical simulations plotted in Fig.~\ref{fig:short_time}(a-b) show how the transition from the local to the nonlocal regime takes place.
Discrepancies first arise in correspondence to the peak and the tails of the field, where $|\nabla^2 |\bm{E}_t|^2|$ is larger. The net effect is smoothing out the peak of the
perturbation, with tails extending in time until reaching the closest boundary \cite{Alberucci:2007}. Numerical simulations confirm that the temporal dynamics of the maximum
reorientation angle $\psi_m$ (i.e., $\psi_m(t)/\psi_\infty$, where $\psi_\infty=\lim_{t\rightarrow\infty}\psi_m(t)$ is the reorientation in the stationary regime) is independent
from the input power as long as the small perturbation condition $\psi\ll\theta_0$ holds true. For example, for $\theta_0=\pi/4$ and $w=6~\mu$m, up to $P=5~$mW no appreciable variations occur in the curves $\psi_m(t)/\psi_\infty$ for different powers, whereas at $P=10~$mW a maximum discrepancy of 2.8$\%$ is observed.\\
Above (see results plotted in Fig.~\ref{fig:short_time}) we considered a fixed optical excitation. The next step is to account for the influence of the molecular rotation on light
propagation \cite{Bloisi:1988}, accounted by Eq.~(\ref{eq:NLSE}), but conserving the $z$-invariance of our system, similarly to the approach employed in
Ref.~\cite{Alberucci:2014}. We thus consider the average value of the beam across $z$, neglecting diffraction or soliton breathing \cite{Alberucci:2015_2}. Specifically, we
approximate the nonlinear index well with a cylindrically-symmetric parabolic index well, in turn providing a Gaussian profile $I =\frac{2P}{\pi w^2(t)} \exp{\left[-\frac{2r^2}{w^2(t)}\right]}$ for the beam,  where $w(t)$ is now time-dependent due to the self-focusing. We thus make the approximation $\Delta n_e^2 \approx 2 n_e(\theta_b) \left.\frac{dn_e}{d\theta}\right|_{\theta_b} \psi_2 r^2$. The coefficient $\psi_2$ is computed by a best-fitting procedure for $\psi$ in the interval $x\in[-\sqrt{2}w(t),\ \sqrt{2}w(t)]$ \cite{Ouyang:2006}. 
\begin{figure}[htbp]
\centering
\includegraphics[width=0.47\textwidth]{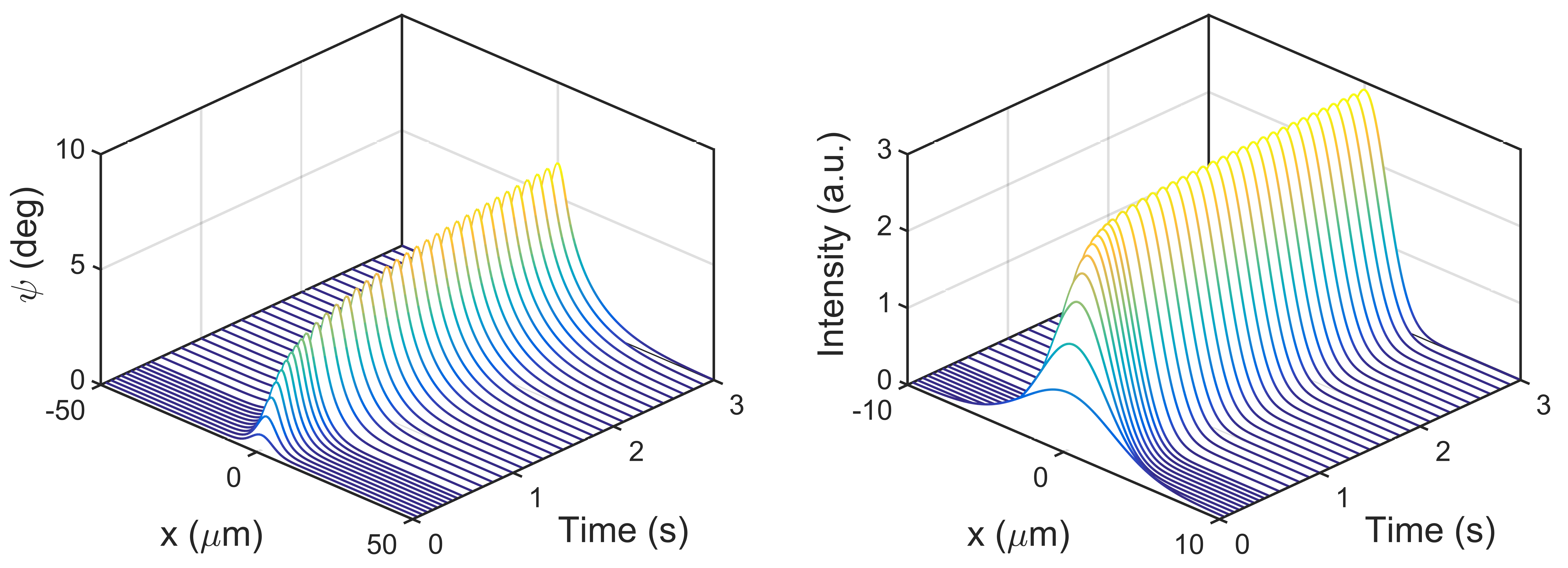}
\caption{Cross-section along the symmetry axis $y=0$ of the optically-induced rotation angle $\psi$ (left side) and of the soliton intensity profile (right side), versus $x$ and $t$. Input
power $P$ and input beam width $w_0$ are 5~mW and $6~\mu$m, respectively.
}
\label{fig:profiles_vs_x_time}
\end{figure}
To calculate the temporal dynamics of the strongly coupled light-matter system, we use the following method. At the initial time $t=0$, we consider an impinging Gaussian beam of waist
$w_0=w(t=0)$. We then compute one temporal step of Eq.~(\ref{eq:reorientation}). From the knowledge of the new rotation angle $\theta$, we can compute $\psi_2$, the latter being necessary to compute the new beam width. This basic procedure is iterated on the temporal interval of interest. Hence, at every step we are able to account for the effect of self-focusing on the intensity distribution. Finally, in the numerical algorithm we set the constraint $w(t)\leq w_0$ so that the beam width cannot be larger than the initial value. Figure~\ref{fig:profiles_vs_x_time} shows the results for $P=5~$mW and $w_0=6~\mu$m. In agreement with
Eq.~(\ref{eq:validity_short_time}), the optical perturbation $\psi$ is local at the beginning (few tens of milliseconds), whereas diffusion comes into play at later times. After the stationary regime is achieved, the profile of the reorientation angle is dictated by the closest boundary (see Section~\ref{sec:stationary_regime}) \cite{Alberucci:2010_2,Minovich:2007}. On the other hand, the beam profile in the first milliseconds is not changing
because the predicted soliton would be wider than the input beam.

\begin{figure}[htbp]
\centering
\includegraphics[width=0.47\textwidth]{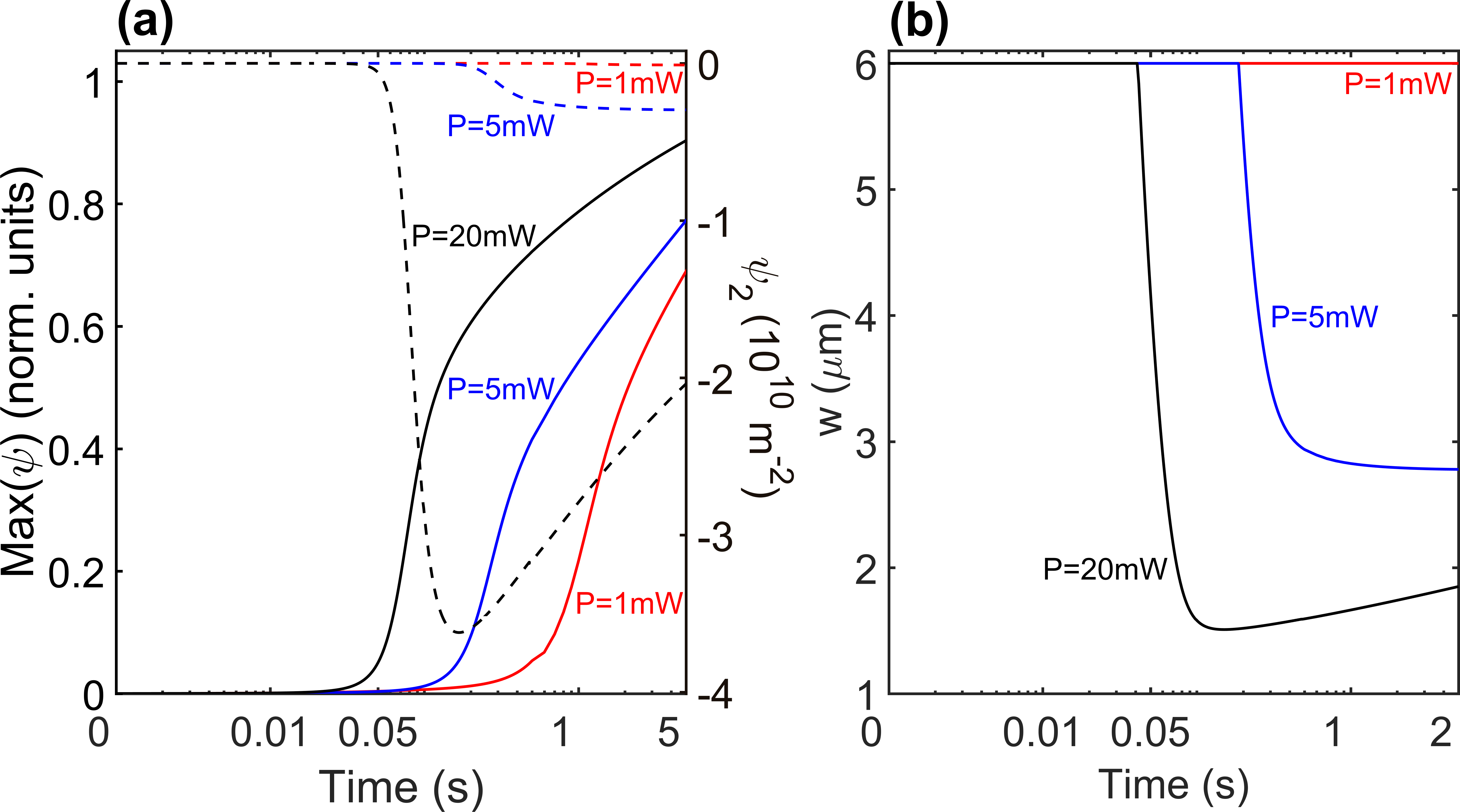}
\caption{Optical behavior in the presence of self-focusing. (a) Maximum of the optical perturbation $\psi_m$ normalized with respect to the stationary value $\psi_\infty$ (left axis, solid lines) and effective quantum harmonic oscillator
strength $\psi_2$ (right axis, dashed lines) versus time. (b) The corresponding beam width versus time. In (a-b) the power is 1~mW (red curves), 5~mW (blue curves) and 20~mW (black curves),
whereas $w_0=6~\mu$m is kept constant.
}
\label{fig:reor_vs_time}
\end{figure}

\begin{figure}[htbp]
\centering
\includegraphics[width=0.47\textwidth]{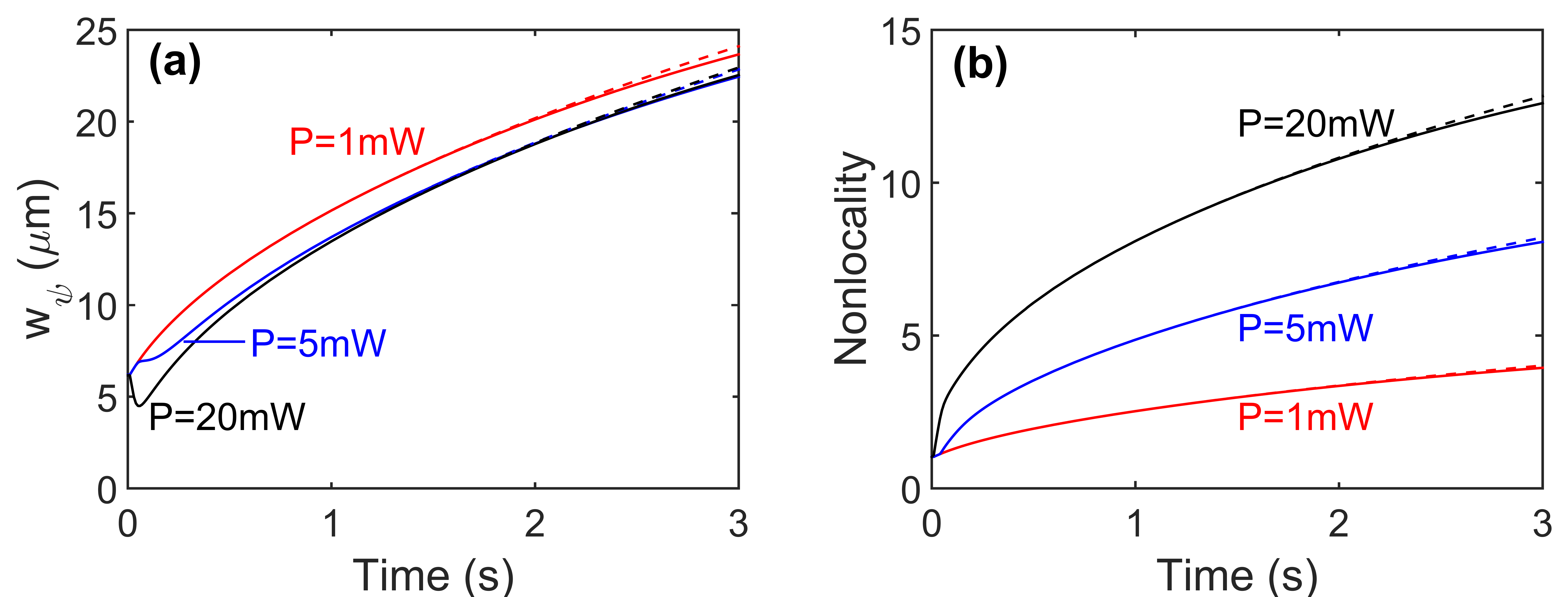}
\caption{(a) Width of the nonlinear perturbation $w_\psi$ and (b) the corresponding nonlocality $w_\psi/w$ versus time, computed across $x$ (solid lines) and $y$ (dashed lines). In
(a-b) the power is 1~mW (red curves), 5~mW (blue curves) and 20~mW (black curves), whereas $w_0=6~\mu$m is kept constant.}
\label{fig:nonlocality_vs_time}
\end{figure}

Figure~\ref{fig:reor_vs_time} shows in more detail the dynamics for different input powers.
When the self-focusing kicks in, the beam starts to shrink towards the stationary soliton solution (for $P=5~$mW it is $w_s\approx 2.8~\mu$m).
Noteworthy, the beam reaches its final size after about 1~s [Fig.~\ref{fig:reor_vs_time}(b)], whereas the maximum and the width of the reorientation angle $\psi$ keep increasing on a longer temporal interval [Fig.~\ref{fig:reor_vs_time}(a), Fig.~\ref{fig:nonlocality_vs_time}(a)].
For $P=1~$mW, the soliton width is larger than $w_0$, the latter meaning no changes in the
optical beam according to our simplified model. Thus, the reorientation follows the case for fixed beam previously investigated (see Fig.~\ref{fig:short_time}). For $P=5~$mW, self-focusing takes place given that the soliton width in the stationary regime is $2.8~\mu$m. Beam narrowing starts after 42~ms, leading to a fast decrease in the beam width. After about
1~s, the nonlinear confinement parameter $\psi_2$ (thus, the beam width $w$) reaches the stationary value. Despite that, the maximum rotation angle keeps increasing (achieving $95\%$
of the stationary value at $t\approx 20~$s), but without appreciable effects on the self-trapped beam. Reorientation in time is slightly steeper for $P=5~$mW than for
$P=1~$mW due to the narrower beam, leading to a faster reorientation [see Eq.~(\ref{eq:sol_transf})].  When $P=20~$mW, both reorientation and beam focusing dynamics are steeper than
for $P=5~$mW (now the $95\%$ of the stationary value is achieved at $t\approx 12~$s). Self-focusing now starts after 8~ms. Due to the fact that $\psi$ is now comparable
with $\theta_0$ (i.e., we are in the nonperturbative regime), the shape of the normalized reorientation curve $\psi_m/\psi_\infty$ now strongly differs from the two previous cases.
The nonperturbative behavior of the nonlinearity is responsible for a non-monotonic behavior in time for $\psi_2$ and $w$, as well. Summarizing, the self-focusing accelerates the
reorientation dynamics owing to a stronger external torque associated with a narrower optical beam, in agreement with Eq.~(\ref{eq:validity_short_time}).\\
In Fig.~\ref{fig:nonlocality_vs_time} we quantitatively analyze the transition from a local to a nonlocal response with time. Figure~\ref{fig:nonlocality_vs_time}(a) shows the width of the optically-induced perturbation $\psi$, defined through the second central moment as $w_\psi=2\sqrt{\int \psi \eta^2 d\eta/\int \psi d\eta}\ (\eta=x,y)$ with respect to time.
The perturbation width is monotonically increasing both
for $P=1$ and 5~mW, whereas for 20~mW an initial shrinking is observed due to the stronger amount of self-focusing. In the stationary regime $w_\psi$ is the same for input powers
larger than few mWs (the shape of the solution of Eq.~(\ref{eq:reorientation}) is almost independent on the soliton width $w$), whereas at $1~$mW the size of the beam leads to a wider reorientation
in space. Although the asymmetric boundary conditions, beam width $w_\psi$ is almost the same along $x$ and $y$, i.e., the nonlinear index well is cylindrically symmetric within a
good accuracy. Figure~\ref{fig:nonlocality_vs_time}(b) shows the temporal evolution of nonlocality $\sigma$ defined as the ratio between the widths of the optical perturbation and
of the intensity profile, $\sigma=w_\psi/w$. At the initial time $t=0$, the medium is local and $\sigma=1$, in agreement with Eq.~(\ref{eq:reorientation_short_time}). Nonlocality
then increases monotonically in time due to the diffusion effect in agreement with Eq.~(\ref{eq:sol_transf}), the larger the power the steeper the increase is.
\begin{figure*}[htbp]
\centering
\includegraphics[width=0.97\textwidth]{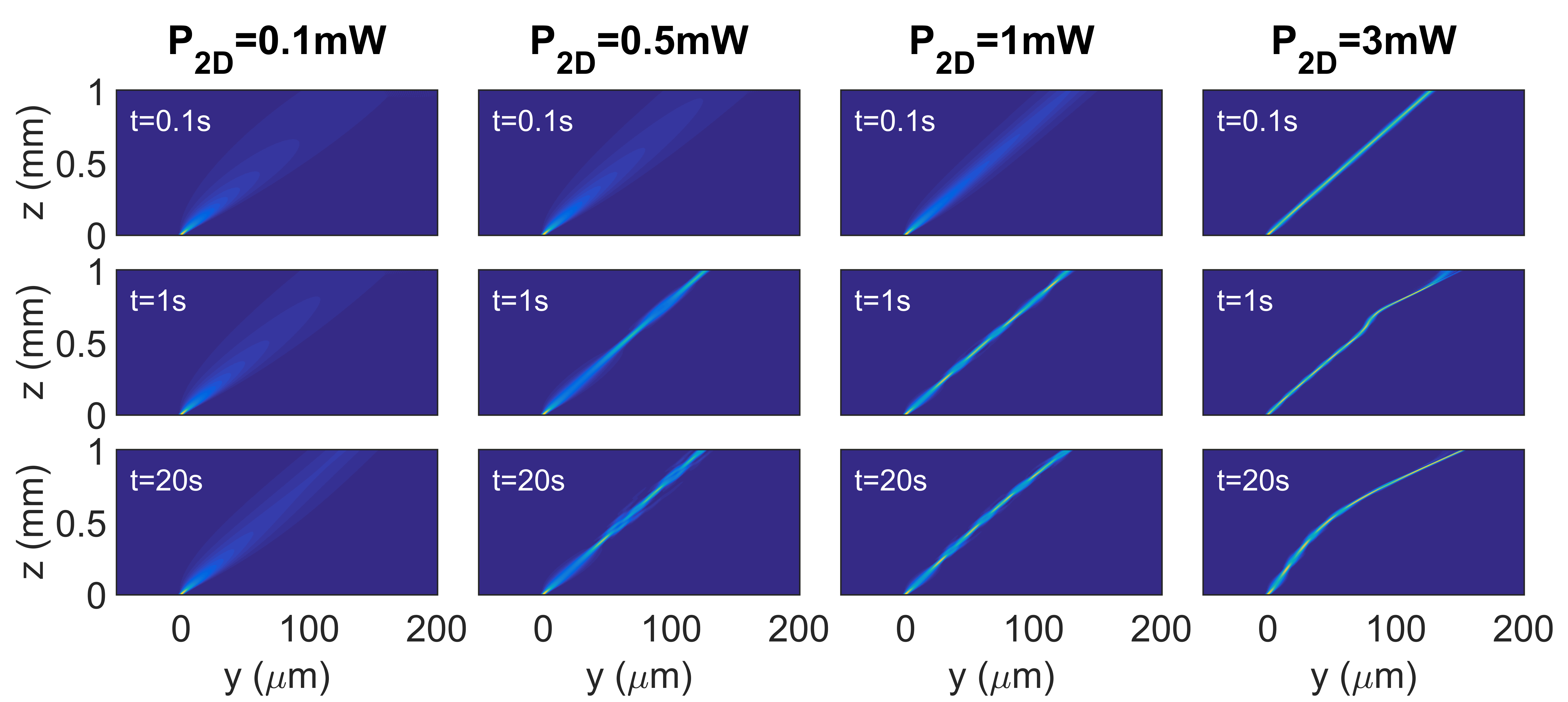}
\caption{Intensity distribution on the plane $yz$ computed via the effective two-dimensional model for $w_\mathrm{in}=2~\mu$m.}
\label{fig:intensities_BPM}
\end{figure*}

\section{Formation of the waveguides: the longitudinal dynamics}
\begin{figure}[h]
\centering
\includegraphics[width=0.47 \textwidth]{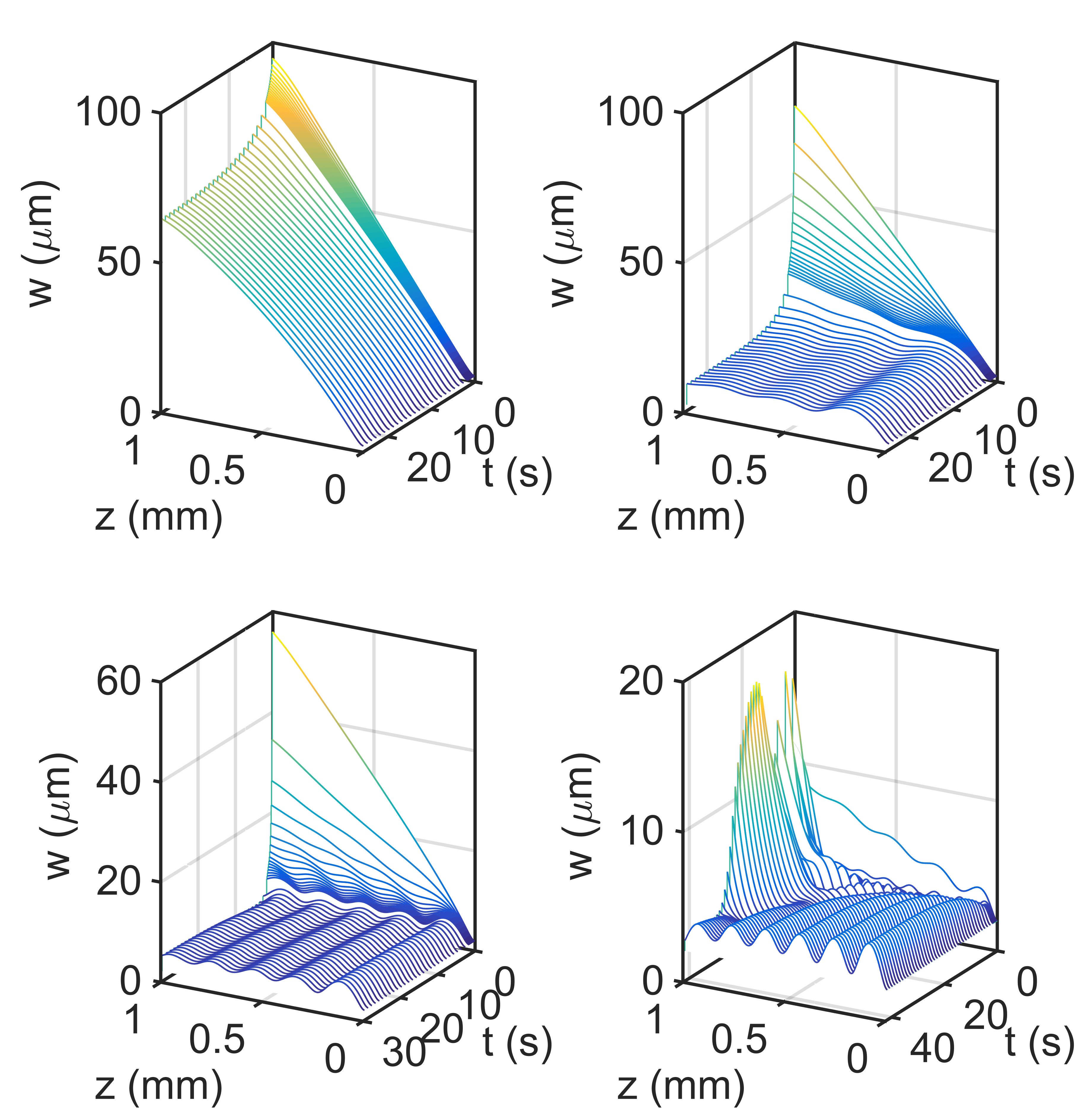}
\caption{Beam width $w$ versus the propagation distance $z$ and time $t$ for $P_{2D}=0.1~$mW (top left), 0.5~mW (top right), 1~mW (bottom left) and 3~mW (bottom right).}
\label{fig:width_BPM}
\end{figure}
In Section~\ref{sec:z_invariant} we described the temporal evolution of light self-trapping neglecting the dynamics of the system along $z$. Such simplified model, despite
providing a qualitative picture of the dynamics, does not account for relevant physical effects, such as self-steering \cite{Alberucci:2010_2}, finite boundary conditions along $z$
\cite{McLaughlin:1995}, oscillation in the beam width \cite{Conti:2004}. The exact 3D numerical solution of Eqs.~(\ref{eq:NLSE}-\ref{eq:reorientation}) is computationally highly
demanding, thus we will account for the longitudinal evolution of the system by using a simplified (1+1)D model (see Section~\ref{sec:effective_model}). In the simplified model we
define an effective power $P_{2D}$, which is about 3$\sim$10 times smaller than the real power $P$ \cite{Alberucci:2010_2}. In this section all the simulations are carried out
considering an input Gaussian beam with planar phase front in $z=0$ and waist $w_\mathrm{in}=2~\mu$m. Finally, the cell length along the propagation distance $z$ is set to $1$~mm.\\
Snapshots of typical light distribution on the plane $yz$ are shown in Fig.~\ref{fig:intensities_BPM}. Videos of the temporal dynamics for $P_{2D}=0.5,~1$ and 3~mW on short (time window of 1~s)  and long scales (tens of seconds) are reported in Visualization 1-3 and 4-6, respectively. For any input power, the amount of light self-localization grows with time.
The effect of nonlinearity on the beam width $w(z,t)$ is plotted in Fig.~\ref{fig:width_BPM}. At low power ($P_{2D}=0.1~$mW), the net effect of nonlinearity is a decrease in the
spreading angle. At $P_{2D}=0.5~$mW, the beam undergoes a strong focusing effect in the first 2~s, whereas for longer times slight adjustments towards the stationary condition occur.
The final condition is a breather, that is, a self-localized mode subject to quasi-periodic oscillations during propagation \cite{Conti:2004,Kivshar:2003}. At $P_{2D}=1~$mW, spatial self-localization occurs faster, but the overall behavior is very close to the case $P_{2D}=0.5~$mW. In all of these cases, the beam width undergoes a continuous transition from a linear spreading (diffraction) to a localized breather. Light propagation for $P_{2D}=3~$mW is more complicated. In fact, in this case the optical perturbation $\psi$ becomes comparable with $\theta_0$, thus light is able to change its own path by changing the walk-off angle \cite{Piccardi:2010}. The effect is already visible at $t=1~$s, where the beam is propagating along a curved trajectory (see Fig.~\ref{fig:intensities_BPM}). A longer temporal interval is necessary before reaching the stationary regime: after $20~$s the light trajectory is not a straight line yet, as it should be in the stationary regime \cite{Alberucci:2010_2}. As a matter of fact, at the beginning the beam oscillates in a seemingly random fashion in time. Then the peaks and valleys on the beam width computed in the stationary regime emerge one by one from the beam wiggling, starting from the input interface up to the output section (last panel in
Fig.~\ref{fig:width_BPM}).\\
\begin{figure}[htbp]
\centering
\includegraphics[width=0.47\textwidth]{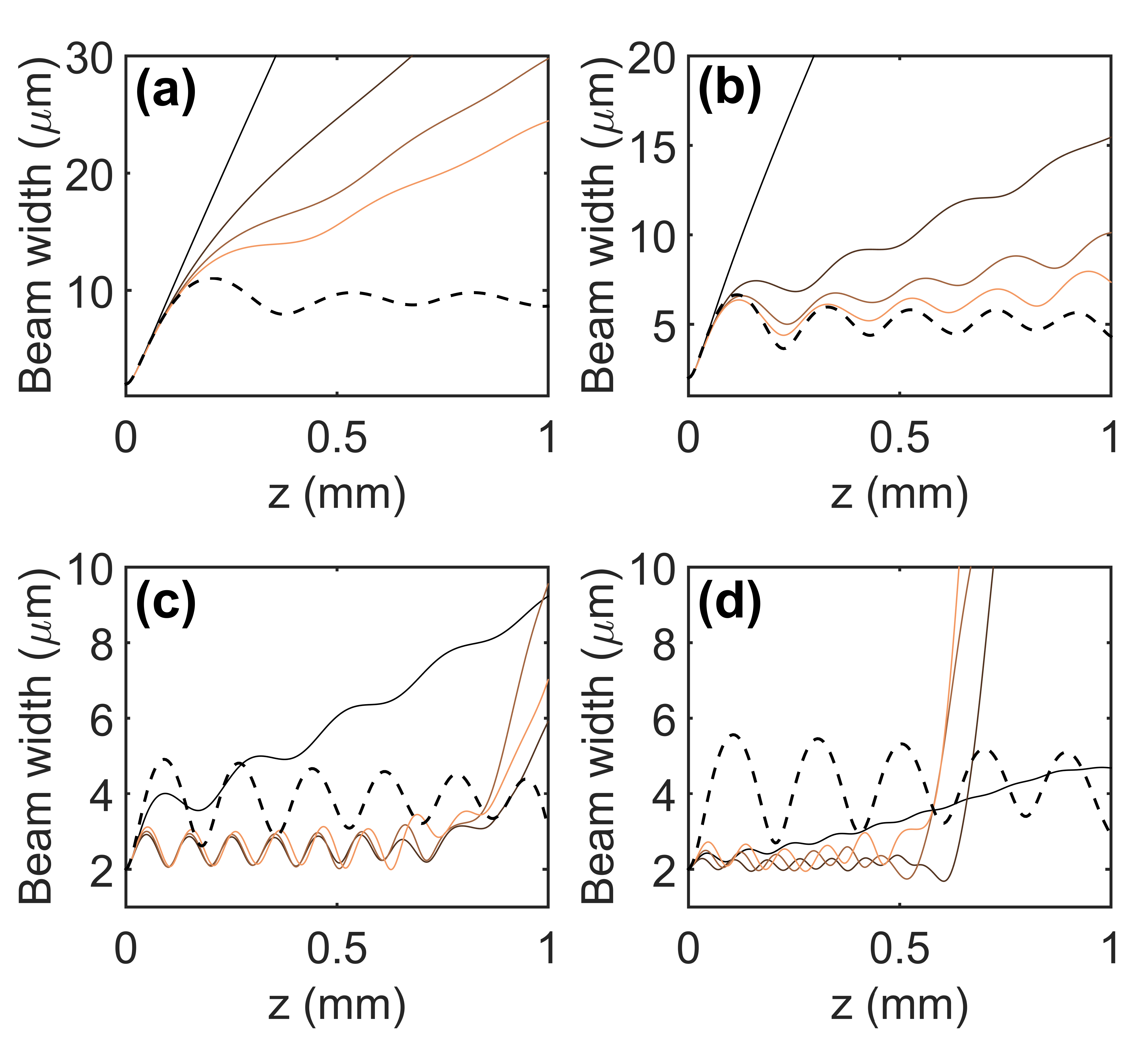}
\caption{Short-time dynamics. Beam width $w$ versus the propagation distance $z$ for $P_{2D}=0.5~$mW (a), 1~mW (b), 3~mW (c) and 5~mW (d). Initial and final time are 0.1~s (darkest curve) and
2.1~s (brightest curves), respectively (the time step is 0.5~s, time growing from the darkest to the brightest curve). Dashed lines corresponds to the stationary solution.}
\label{fig:width_BPM_short_time}
\end{figure}
From an engineering point of view, we are particularly interested in the minimum time required to form the waveguide. To address this point, the short-time dynamics of the beam width is
plotted in Fig.~\ref{fig:width_BPM_short_time} (see also Visualization 1-3). Regardless of the input power, self-confinement takes place first close to the input interface, eventually moving to larger $z$
with time. The time required for the formation of waveguides strongly depends on the input power. We will consider the waveguide formed when the beam width at the
output ($z=1$~mm) is lower than 10$~\mu$m. When $P_{2D}=0.5$~mW [Fig.~\ref{fig:width_BPM_short_time}(a)], the waveguide is not yet formed in the first two seconds, whereas for
$P_{2D}=1$~mW waveguiding is achieved after 1~s [Fig.~\ref{fig:width_BPM_short_time}(b)]. For $P_{2D}>3~$mW, localization takes place at $t=0.1~$s [Fig.~\ref{fig:width_BPM_short_time}(c-d)]. Important to stress, the guides formed in this transient regime differ from the stationary value, the difference increasing with the
input power. In fact, in Fig.~\ref{fig:width_BPM_short_time}(c-d) we notice that for $t>0.5~s$ the guide at large $z$ is destroyed due to trajectory oscillations associated with
self-steering \cite{Braun:1993}.\\
\begin{figure}[htbp]
\centering
\includegraphics[width=0.47\textwidth]{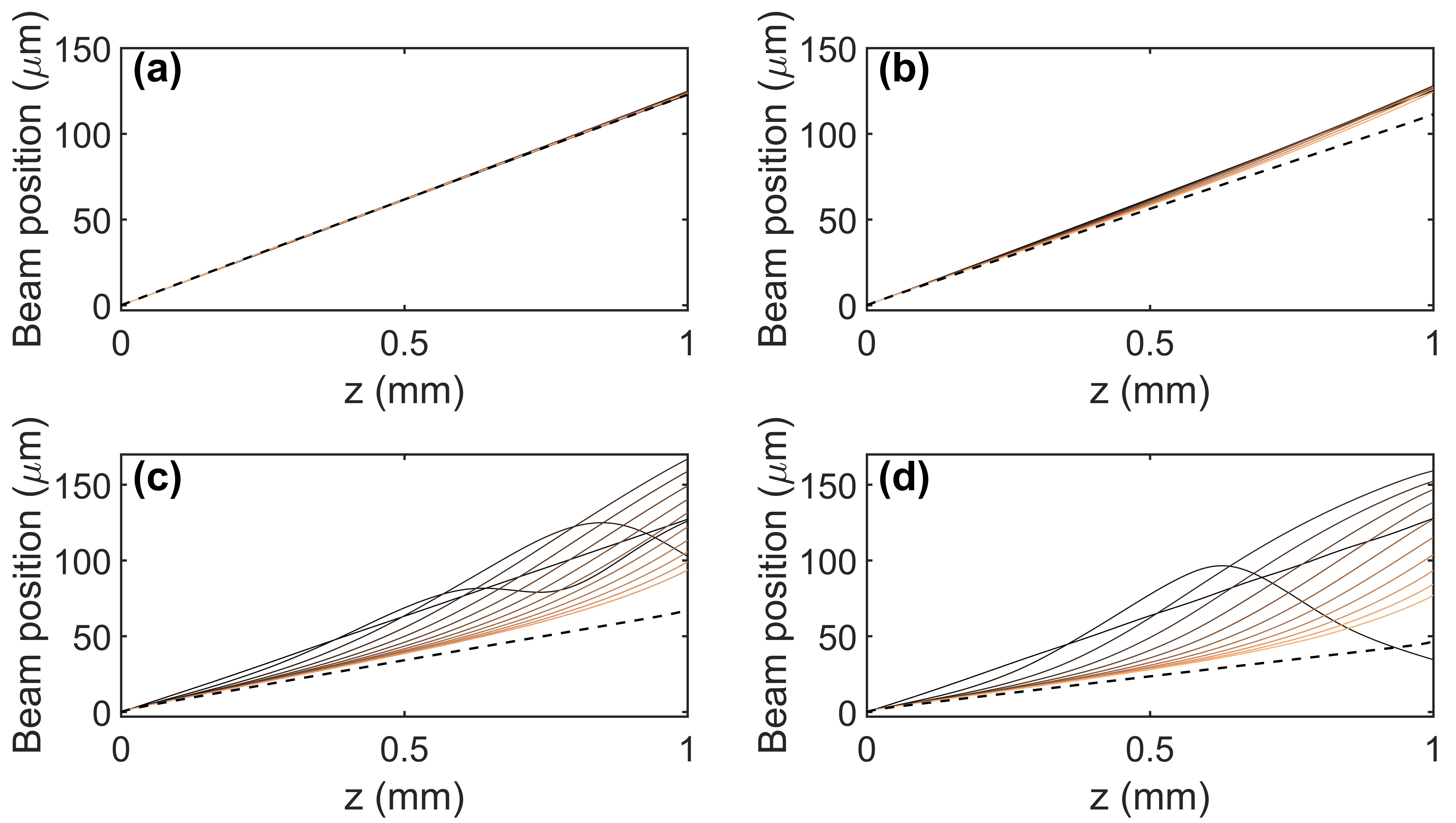}
\caption{Beam trajectory versus the propagation distance $z$ for $P_{2D}=0.5~$mW (a), 1~mW (b), 3~mW (c) and 5~mW (d). Time increases from darkest to brightest curves, respectively,
with a time step of 4~s. Dashed lines corresponds to the stationary solution. }
\label{fig:traj_BPM}
\end{figure}
To summarize, the waveguides can be formed in a temporal interval of hundreds of milliseconds, however, using powers large enough to induce oscillations in the beam path on longer times. The drawback is a longer relaxation time before the stationary regime is reached. Let us then discuss the nature of the oscillations in the beam trajectory (see Visualization 4-6). Figure~\ref{fig:traj_BPM} shows the behavior of the light path, defined as the center-of-mass of the beam. When nonlinear changes in the walk-off are negligible [Fig.~\ref{fig:traj_BPM}(a-b)], the trajectories slowly relax towards the stationary condition. When self-steering is appreciable [Fig.~\ref{fig:traj_BPM}(c-d)], at very short times the beam propagates along straight lines determined by the linear value of the walk-off, $\delta(\theta_0)$. As time goes on, the reorientation angle becomes large enough to misplace the beam with respect to the induced index well, the latter being inhibited to follow instantaneously the changes in the light distribution owing to the slow response of the material. The net effect is that the light beam starts to oscillate around the index well created at the previous time by the light itself. At later times and starting from the input interface, the beam slowly relaxes towards the final trajectory, corresponding to a straight line with a lower slope than the initial one (walk-off $\delta$ encompasses a maximum in proximity of $\theta=45^\circ$).\\
\begin{figure}[htbp]
\centering
\includegraphics[width=0.47\textwidth]{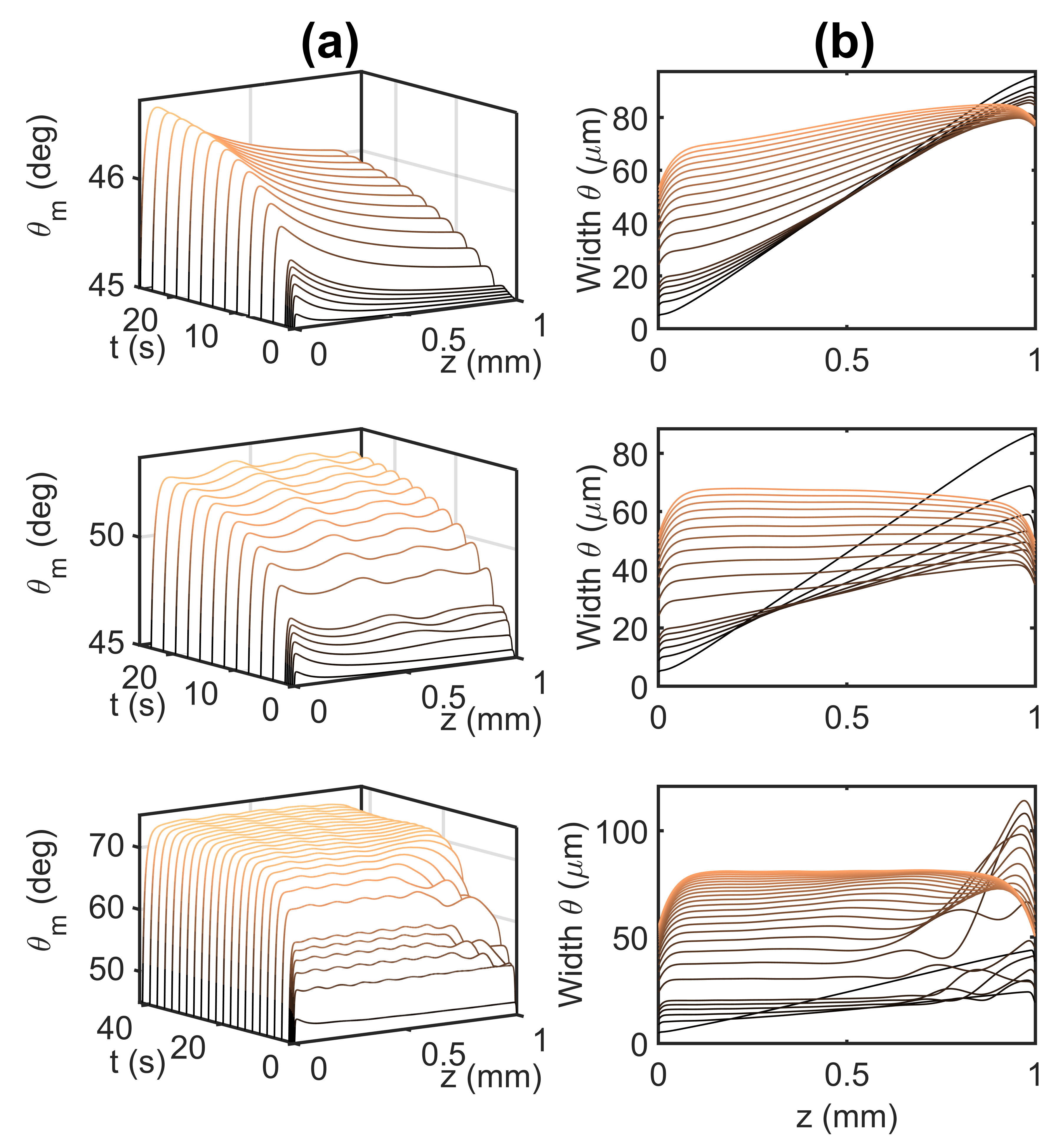}
\caption{(a) Maximum of the rotation angle $\theta_m(z)$ versus time and propagation distance $z$. (b) Width of the optical perturbation $\psi$ versus $z$. Input power is $P_{2D}=0.1~$mW, 0.5~mW, and 3~mW, from top to bottom. Time increases from the darkest to the brightest curves, respectively. Specifically, both in (a) and (b) the time is from 0.1~s to 1.3~s with steps of 0.3~s, and from 1.4~s onwards on steps of 2~s.
}
\label{fig:theta_BPM}
\end{figure}
Next step is to analyze the features of the waveguide and their evolution in time. Figure~\ref{fig:theta_BPM} reports the maximum director angle $\theta$ computed on each plane
$z=\text{const}$ (dubbed $\theta_m(z,t)$) and the width of the waveguide $w_\psi(z,t)$. The magnitude of $\theta_m$ versus time closely resembles ~trends shown in
Fig.~\ref{fig:reor_vs_time}, with larger powers corresponding to faster reorientation. For $P_{2D}=0.5~$mW, $\theta_m$ decays with $z$ owing to the beam spreading in propagation,
see the second panel in Fig.~\ref{fig:width_BPM}. For the same reason, the width of the guide increases with $z$. For larger input powers the maximum reorientation angle is almost
constant along $z$ due to the filtering action of the spatial nonlocality along $z$. When self-steering occurs ($P_{2D}=3$~mW) and in accordance with the oscillation in the beam
path, $\theta$ varies rapidly close to the output interfaces, stabilizing after about 20~s. Regardless of the input power and in agreement with the simplified model discussed in
Section~\ref{sec:z_invariant}, nonlocality increases monotonically with time. Thus, the waveguides observed at short time intervals correspond actually to a different and narrower
spatial profile with respect to the stationary case (see Section~\ref{sec:stationary_regime}).


\section{Discussion of the results and future perspectives}
In this article we investigated the temporal dynamics associated with the formation of light-written waveguides in liquid crystals.  We showed how the degree of spatial nonlocality evolves in time from a local (the waveguide shape is identical to the intensity profile) to a highly nonlocal regime (the shape and extension of the waveguide is dictated by the sample geometry), such result holding valid in thermal materials as well \cite{Rothschild:2006,Kaminer:2007,Navarrete:2017}. An interesting consequence of our results is the possibility to control the waveguide spatial extension varying the repetition rate of a train of pulses. \\
With respect to the reorientational nonlinearity in NLC, two main regimes have been found. When the input power is not large enough to induce self-steering (the perturbative regime), the nonlocal waveguide is formed in few seconds, and the formation time is shorter for increasing powers. Despite that, the full stationary regime is achieved in tens of seconds. The multi stage dynamics arises from the different time constants associated with every spatial mode of the nonlinear perturbation (the wider the mode the longer the response time is). Importantly, these results hold valid in thermal materials such as lead glasses \cite{Dabby:1968, Rotschild:2005, Alfassi:2006} or, more in general, in the presence of a nonlinear mechanism based upon diffusion (we stress that in NLC more complicated thermal effects can appear owing to convective heat exchange or owing to the presence of the nematic-isotropic phase transition \cite{Simoni:1997,Shih:2010}). In the nonperturbative regime, the power is large enough to induce a power-dependent beam trajectory. The nonperturbative regime is typical of the reorientational nonlinearity in NLC, and it can be described only accounting for the power-dependent spatial walk-off and the longitudinal evolution of the self-written waveguide. In the nonperturbative regime, the beam starts to fluctuate during the transient regime owing to the misalignment between the beam and the waveguide. As a matter of fact, the stationary regime is reached after a longer time interval with respect to the perturbative regime. Analogously to the perturbative case, a local waveguide can be written after fractions of seconds.\\
Our results demonstrate that temporal modulation of the impinging beam represents an additional degree of freedom in the control of light-written waveguides in nonlocal nonlinear media \cite{Henninot:2006,Vocke:2017}. In a broader perspective, our results represent the first step towards a full investigation of the interplay between temporal and spatial effects in highly nonlocal materials.
With respect to the field of light self-localization in nematic liquid crystals, future developments include the assessment of the role played by the strong scattering associated with the collective molecular fluctuations, the latter being associated with a temporal instability for self-trapped beam at large enough powers \cite{Braun:1993,Valkov:1985,Bolis:2017}. Finally, in future works it will be interesting to pursue the generalizations of the phenomena discussed here to the case of smectic phase, where much faster time responses with respect to the nematic phase can be achieved, but at the expense of stronger elastic interactions \cite{Clark:1980}.

\textbf{Funding.}  Deutsche Forschungsgemeinschaft (DFG) (GRK 2101); Academy of Finland, Finland Distinguished Professor grant No. 282858; ERC Adv. grant No. 694683 (PHOSPhOR); FCT grant No. SFRH/BPD/77524/2011


\section*{Appendix}

\subsection{Time-dependent solution of a forced diffusion equation}
\label{sec:diffusion_time}
Let us consider the normalized equation
\begin{equation}  \label{eq:diffusion_normalized}
 \frac{\partial f(\bm{r},t)}{\partial t}=D \nabla^2 f(\bm{r},t) + F(\bm{r},t),
\end{equation}
where $F(\bm{r},t)$ is the forcing term, in the most general case dependent on time as well.
Using the 3D spatial Fourier transform, $\tilde{f}(\bm{k},t)=(2\pi)^{-\frac{3}{2}}\iiint{f(\bm{r},t) e^{i\bm{k}\cdot\bm{r}} d^3\bm{r}}$, we find the solution in the transformed domain
\begin{equation}  \label{eq:sol_transf_time}
  \tilde{f}(\bm{k},t) = \tilde{f}(\bm{k},0)e^{-\bm{k}^2 D t} + \int_0^t{e^{-\bm{k}^2 D(t-t^\prime)}\tilde{F}(\bm{k},t^\prime)dt^\prime}.
\end{equation}
If $F$ is independent on time, we get
\begin{equation} \label{eq:sol_transf}
 \tilde{f}(\bm{k},t)= \tilde{f}(\bm{k},0)e^{-\bm{k}^2 Dt} + \frac{\tilde{F}(\bm{k})}{\bm{k}^2 D}\left(1- e^{-\bm{k}^2 D t}\right).
\end{equation}
In the stationary regime, $t\rightarrow\infty$
\begin{equation} \label{eq:sol_stationary}
 \tilde{f}_{stationary}(\bm{k})=  \frac{\tilde{F}(\bm{k})}{\bm{k}^2 D},
\end{equation}
corresponding to the solution of the Poisson equation. Boundary conditions, i.e., the shape of the sample, affect Eq.~(\ref{eq:sol_stationary}) by selecting the allowed spatial
frequencies $\bm{k}$. The stationary solution provided by Eq.~(\ref{eq:sol_stationary}) does not depend on the initial condition $f(\bm{r},0)$. For short time intervals,
Eq.~(\ref{eq:sol_transf}) yields
\begin{equation}  \label{eq:transf_short_time}
  \tilde{f}(\bm{k},t)= \left(1-\bm{k}^2 D t\right) \tilde{f}(\bm{k},0) + {\tilde{F}(\bm{k})}t,
\end{equation}
corresponding in the real domain to
\begin{equation}
  f(\bm{r},t)= f(\bm{r},0) + \left( D \nabla^2 f(\bm{r},0)+ F(\bm{r}) \right) t.
\end{equation}

\subsection{Shape-preserving solitons in the stationary regime}
\label{sec:stationary_regime}
\begin{figure}[htbp]
\centering
\includegraphics[width=0.47\textwidth]{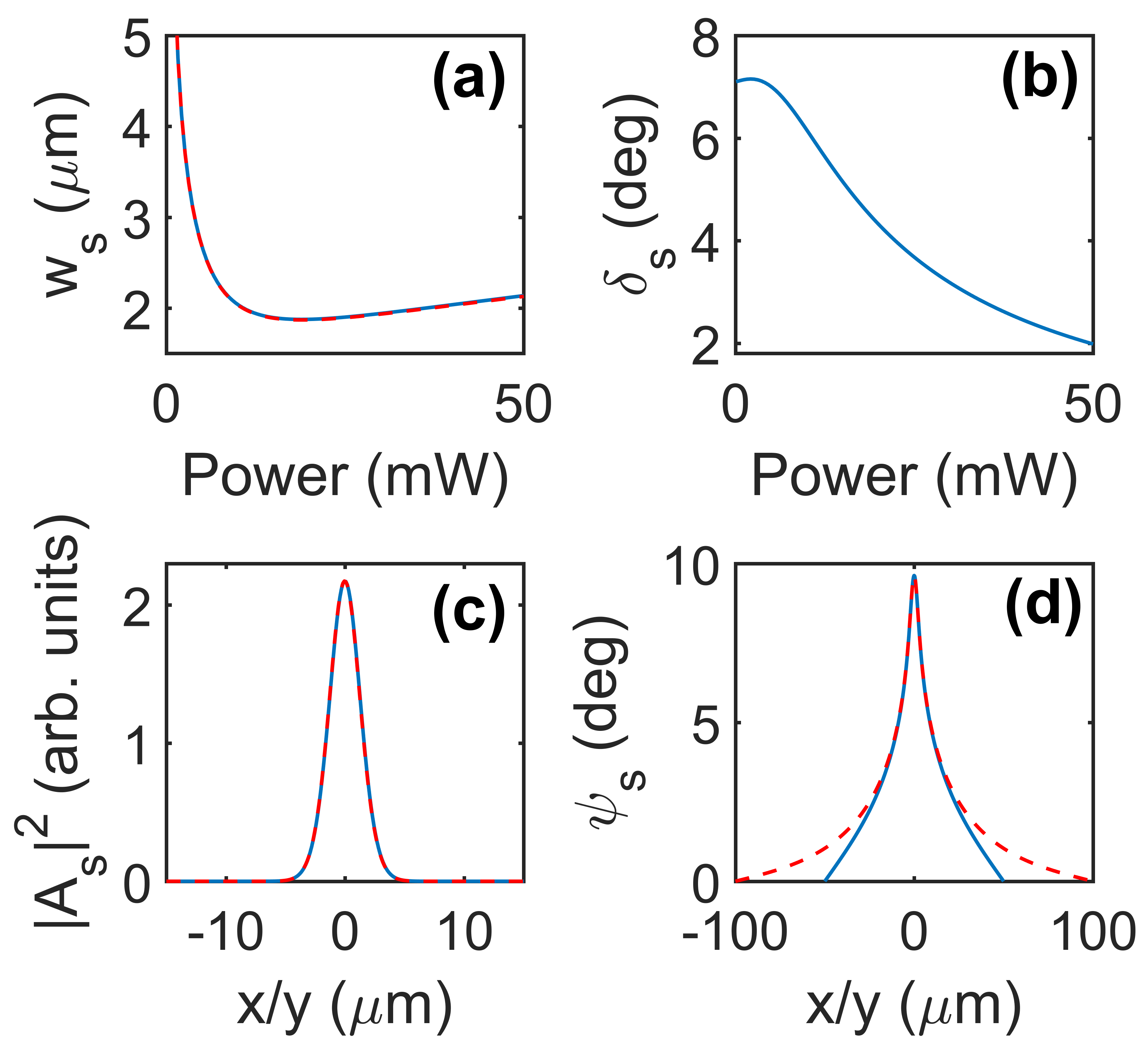}
\caption{(a) Beam width versus input power along $x$ (solid blue line) and $y$ (red dashed line) measured by best-fitting with an astigmatic Gaussian function. (b) Walk-off angle $\delta_s$
versus the soliton power. Cross-sections of the soliton intensity profile (c) and reorientation angle (d) versus $x$ with $y=0$ (blue solid line), and versus $y$  with $x=0$ (red dashed line). }
\label{fig:soliton}
\end{figure}
The fundamental shape-preserving solitons of Eqs.~(\ref{eq:NLSE}-\ref{eq:reorientation}) in the stationary regime (i.e., $\partial_t \theta=0$) are found by setting $A=A_s(x,y-\tan\delta_s z)e^{i\beta_s z}$ and $\theta=\theta_s(x,y-\tan\delta_s z)$. The ansatz corresponds to a soliton with planar phase-front and propagating along a walk-off angle
$\delta_s$, the latter being also dependent on the input power in the non-perturbative regime. Then Eqs.~(\ref{eq:NLSE}-\ref{eq:reorientation}) turn into a nonlinear eigenvalue
problem parametrized with respect to the input power $P$. The eigenfunction $A_s$ and the eigenvalue $\beta_s$ of the electromagnetic problem are found as the corresponding
eigenvector and eigenvalue of the discrete operator corresponding to Eq.~(\ref{eq:NLSE}).
The optical profile found is then substituted into Eq.~(\ref{eq:reorientation}) and the new director rotation is obtained. The new material parameters are then substituted back into Eq.~(\ref{eq:NLSE}), and the procedure is iterated until convergence is achieved.\\
Figure~\ref{fig:soliton}(a) and (b) show the beam width and the walk-off angle $ \delta_s$ versus power $P$. The beam width is not monotonically shrinking owing to the saturation
effect related to large reorientation angles, in turn weakening the effective nonlinearity. Figure \ref{fig:soliton}(c) and (d) show respectively the functions $A_s$ and
$\theta_s$ versus the transverse coordinates for $P=5.1~$mW. Shapes of the soliton changes slightly with power, conserving a quasi-Gaussian profile. The reorientation also maintains
its transverse profile, the latter being determined mainly by the boundary conditions along $x$, i.e., the shortest size of the sample \cite{Hutsebaut:2005,Minovich:2007}. Despite
the asymmetric boundary conditions, the reorientation angle in proximity to the cell center is cylindrically symmetric.

\subsection{Effective (1+1)D model}
\label{sec:effective_model}

Numerical simulations of Eqs.~(\ref{eq:NLSE}-\ref{eq:reorientation}) are highly demanding from the computational time point of view. An effective bidimensional problem can be solved
in behalf of Eqs.~(\ref{eq:NLSE}-\ref{eq:reorientation}). The model is \cite{Alberucci:2010_2}
\begin{align}
 & 2i k_0 n_e(\theta_b) \left(\frac{\partial A}{\partial z} +\tan\delta(\theta_b) \frac{\partial A}{\partial y} \right) + D_y \frac{\partial ^2 A}{\partial y^{ 2}} \nonumber \\ &+ k_0^2 \Delta n_e^2(\theta) A=0. \label{eq:NLSE_2D} \\
 \nu \frac{\partial \psi}{\partial t} &= K\left(\frac{\partial^2 \psi}{\partial y^2} + \frac{\partial^2 \psi}{\partial z^2} \right) - K\left(\frac{\pi}{L_x}\right)^2 \psi \nonumber \\  &+ \frac{\epsilon_0 \epsilon_a}{4} \sin\left[2\left(\theta_0+\psi-\delta_b  \right) \right] \left( |\bm{E}_t|^2 - |\bm{E}_s|^2 \right) \nonumber \\ & +  \frac{\epsilon_0 \epsilon_a}{2} \cos\left[2\left(\theta_0+\psi-\delta_b \right) \right] \text{Re}\left(\bm{E}_t \bm{E}_s^* \right),
    \label{eq:reorientation_2D}
\end{align}
with an effective power $P_{2D}$ lower than the real one. The screening term in Eq.~(\ref{eq:reorientation_2D}) is inserted in order to conserve the nonlocality degree of the
original system \cite{Kaminer:2007}. According to the results shown in Section~\ref{sec:diffusion_time}, the response time depends critically on the shape of the Green function for
the material response. Thus, the simplified model conserves the response time of the original 3D model.

\bibliography{references}

\end{document}